\documentclass[aps,prc,twocolumn,endfloats,floatfix,showpacs]{revtex4}
\usepackage{amssymb}
\usepackage{amsmath}
\usepackage{graphicx}
\usepackage{dcolumn}
\usepackage{color}

\input{tcilatex}

\begin{document}

\preprint{HEP/123-qed}
\title{Connection between the "Strutinsky level density" and the
semiclassical level density}
\author{B. Mohammed-Azizi}
\email{aziziyoucef@voila.fr}
\affiliation{University of Bechar, Bechar, Algeria}
\author{D. E. Medjadi}
\email{demedjadi@voila.fr}
\affiliation{Ecole Normale Superieure, Kouba, Algiers, Algeria}
\date{\today}

\begin{abstract}
We establish an analytical link between the level density obtained by means
of the Strutinsky averaging method, and the semiclassical level density.
This link occurs only in the so-called "asymptotic limit".

It turns out that the Strutinsky method amounts to an approximation to the
semiclassical method. This approximation contains an unavoidable remainder
which constitutes an intrinsic noise in comparison to the semiclassical
method. Thus, the "old" problem of the dependency of the Strutinsky
procedure on the two free smoothing parameters of the averaging is
intimately connected to this noise.

On the other hand, we demonstrate that the noise of the method is small in
the average density of states and in the average energy, whereas it might be
non-negligible in the shell correction itself. In order to improve this
method, we give a "rule" which consists simply of minimizing the relative
error made on the average energy.
\end{abstract}

\keywords{Nuclear physics, Nuclear structure, level density, Strutinsky
averaging method, Wigner-Kirkwood expansion}
\pacs{21.10Dr, 21.10.Ma, 21.60.-n}
\volumeyear{year}
\volumenumber{number}
\issuenumber{number}
\eid{identifier}
\date[Date text]{date}
\received[Received text]{date}
\revised[Revised text]{date}
\accepted[Accepted text]{date}
\published[Published text]{date}
\startpage{1}
\endpage{}
\maketitle

\section{Introduction}

The inclusion of the Strutinsky's shell correction \cite{1}-\cite{3} in the
liquid- drop model \cite{4}, namely the so-called macroscopic-microscopic
method, has allowed considerable improvements in the predictions of the
nuclear masses \cite{5}, and in the calculations of the fission barriers, as
well \cite{6},\cite{8}. Nowadays, despite the progress of the more basic
microscopic models (such as self-consistent models), it remains often used.

This method consists essentially of combining the liquid-drop model
(macroscopic model) where the binding energy varies slowly as a function of
the number of nucleons $N$ and $Z$, with a shell correction varying abruptly
with $N$ and $Z$. The latter is due to the non-uniformity of the shell
structure of the energy levels. It is extracted from a single-particle
Hamiltonian (microscopic model) according to an original idea of Strutinsky.

The Strutinsky method is mainly based on a particular smoothing procedure of
the density of states.

Although this method is very efficient, it contains two weak points which
are:

1) The dependence of the results on the two well known inherent parameters,
i.e. the width $\gamma $ and the order $M$ of the smoothing

2) The difficulties of the treatment of the continuum encountered with
realistic mean potentials

The purpose of the present work is summarized in the following points:

1) The Strutinsky method can be derived rigorously from the point of view of
the least-squares approximation\emph{\ }of the level density. The
equivalence between this point of view and the well-known standard averaging
appears trivial.

2) In this work, it is proved analytically that the averaged level density
obtained by the Strutinsky method is simply an approximation to the
semiclassical level density.\newline
In this respect, the semiclassical level density can be considered as the
"true" (i.e. exact) smooth density.

3) It is mainly shown that in comparison with the semiclassical method, the
Strutinsky method is characterized by a remainder which contains all the
dependence on the two smoothing (free) parameters, and which is thence, the
source of the "noise" of the averaging procedure.

4) Concerning the smooth density of states and the smooth energy, it is
demonstrated that the Strutinsky method is reliable. However, the shell
correction itself must be treated with care because it is very sensitive to
the choice of the two free averaging parameters. In this context, in order
to improve the method, we propose the "rule of the relative remainder" (RRR
rule).

5) At last, it is explained why the Strutinsky method fails near the
zero-energy (top of the well) for finite mean potentials

\section{The Strutinsky averaging}

\subsection{Bases and phenomenology of the Strutinsky's method}

In spite of the complexity of the nuclear forces, it appears that most of
the binding energy of the nuclei is well described by the so-called
liquid-drop model. This simple phenomenological approach is of a classical
type. This means that the quantum effects, or a more precisely, the shell
effects, are ignored by this model. This causes systematic discrepancies
between the theoretical predictions and the experimental data \newline
On the other hand, it is known that such effects are contained in the shell
model, but the latter is unable to reproduce correctly the general trends of
the binding energy. To solve this dilemma, Strutinsky has proposed to
combine the binding energy of the liquid-drop model with a small (but
essential) correction deduced from the shell model. This can be written as:

\begin{eqnarray*}
E\text{ (Binding Energy)} &\text{=}&E\text{ (Liquid Drop Model)} \\
&&+\delta E\text{ (Shell Correction)}
\end{eqnarray*}

The shell correction is calculated from a mathematical prescription outlined
by Strutinsky

It is obtained by summing the single particle energies of a phenomenological
shell model potential and subtracting the average (smooth) part of this
quantity:

\begin{equation*}
\delta E\text{ }(\text{Shell Correction})=\dsum \limits_{i}\epsilon_{i}-%
\overline{\dsum \limits_{i}\epsilon_{i}}
\end{equation*}

As already mentioned, this method is often called the
macroscopic-microscopic method because it mixes two very different models.
Such a duality is obviously not free from inconsistency. Nevertheless, it is
possible to give a microscopic basis to this "model" within the Hartree-Fock
(HF) approximation and some simple assumptions \cite{2},\cite{3}.\newline
This consists essentially of expanding the HF energy around its
semiclassical approximation obtaining thus the so-called "Strutinsky energy
theorem":

\begin{equation}
E(\rho )=E(\overline{\rho })+\left( \dsum\limits_{i}\epsilon _{i}-\overline{%
\dsum\limits_{i}\epsilon _{i}}\right) +O_{2}  \label{591}
\end{equation}

Here $\rho$ is the HF density matrix, and $\overline{\rho}$ its
semiclassical approximation which is a smooth quantity, free of shell
effects. For this reason $\overline{\rho}$ can be assimilated to the
classical average (i.e. without quantal variations) part of $\rho$. The sums
of single-particle energies $\Sigma\epsilon_{i}$ and $\overline{%
\Sigma\epsilon_{i}}$ are related respectively to $\rho$ and $\overline{\rho}$%
. Finally, $O_{2}$ is a quantity of the second order in the operator $\rho-%
\overline{\rho}$, and is generally negligible (for details see \cite{27}).

It is clear from (\ref{591}) that the macroscopic-microscopic method
described above is a "schematic" interpretation of this theorem. The
quantity $E(\overline{\rho})$ can be replaced by the energy of the
liquid-drop formula, and the shell correction is deduced from a
phenomenological one body Hamiltonian \newline
It is also to be noted that a complete microscopic approach of this theorem
remains possible. For example, in ref. \cite{26} the authors use a method
referred to as the Extended Thomas Fermi plus Strutinsky integral (ETFSI)
method. In the latter, the semiclassical quantity $E(\overline{\rho})$ is
deduced self-consistently from a microscopic effective interaction.(Skyrme
III). The shell correction is then added to $E(\overline {\rho})$.

In the following, we will mainly focus on two points:

\begin{itemize}
\item The mathematical aspect of the Strutinsky smoothing

\item The link between this smoothing and the semiclassical approximation,
and its consequences
\end{itemize}

\subsection{The "exact" level density}

The Strutinsky averaging can be derived by various formal approaches \cite{1}%
-\cite{3}, \cite{6}, \cite{8}-\cite{10}.\newline
In this work, we will derive it from the point of view of the least-squares
approximation. In fact, this point has been early suggested by Bunatian and
co-authors \cite{18}. The least-squares approximation will help us to
understand why the Strutinsky averaging fails near the edge of realistic
potentials.$\newline
$For an entirely discrete spectrum the level density of states is defined by:

\begin{equation}
g_{o}(\epsilon)=\dsum \limits_{n=0}^{\infty}\delta(\epsilon-\epsilon_{n})
\label{1}
\end{equation}
In the following, this density will be called the "exact quantum level
density", or shortly, the "exact level density", because it is a true
quantum quantity, as opposed to its semiclassical approximation, or as
opposed to the level density obtained from $g_{o}(\epsilon)$ by the
Strutinsky averaging (see below).

In fact, Eq.(\ref{1}) concerns uniquely infinite potentials without
continuum. Finite potentials will be treated separately at the end of this
paper.

\subsection{Polynomial approximation to the "exact" level density:}

Let $g_{M}(\epsilon )$ be a polynomial approximation of order $M$ to the
"exact" level density. More precisely, we seek this approximation in the
vicinity of a point $\lambda $ (which represents actually the Fermi level)
in an effective interval $\left[ -\gamma +\lambda ,\lambda +\gamma \right] $%
, by using the Gaussian weight $\exp \left( -(\epsilon -\lambda )^{2}/\gamma
^{2}\right) $. For this reason, the cited polynomial must depend a priori
not only on $M$, but also on $\gamma $ and $\lambda $. Therefore, it will be
denoted $\overline{g}_{M\text{ , }\gamma }(\epsilon ,\lambda )$. For our
purpose, it will be convenient to write this polynomial as a linear
combination of Hermite polynomials $H_{k}$.%
\begin{equation*}
\overline{g}_{M\text{ , }\gamma }(\epsilon ,\lambda
)=\sum\limits_{k=0}^{M}c_{k}H_{k}\left( \dfrac{\epsilon -\lambda }{\gamma }%
\right)
\end{equation*}

Thus, We must look for the polynomial $g_{M\text{ , }\gamma}(\epsilon
,\lambda)$ which minimizes the integral:

\begin{equation}
I(\lambda,M,\gamma)=\int\limits_{-\infty}^{\infty}\left[ g_{o}(\epsilon )-%
\overline{g}_{M\text{ , }\gamma}(\epsilon,\lambda)\right] ^{2}e^{-\left( 
\dfrac{\epsilon-\lambda}{\gamma}\right) ^{2}}d\epsilon  \label{2}
\end{equation}

This procedure is a local averaging in the sense of the least- squares fit.
Minimizing (\ref{2}) with respect to the coefficients $c_{k}$ , and using
the orthogonality property of the Hermite polynomials we find:

\begin{gather}
\overline{g}_{M\text{ , }\gamma }(\epsilon ,\lambda
)=\sum\limits_{m=0}^{M}c_{m}(\lambda ,\gamma )H_{m}(\dfrac{\epsilon -\lambda 
}{\gamma })  \label{33} \\
c_{m}(\lambda ,\gamma )=\dfrac{1}{m!2^{m}\sqrt{\pi }}\dsum\limits_{n=0}^{%
\infty }H_{m}(u_{n})\dfrac{1}{\gamma }\exp \left[ -u_{n}^{2}\right] ,  \notag
\\
u_{n}=\dfrac{\epsilon _{n}-\lambda }{\gamma }  \notag
\end{gather}

As we shall see in the next subsection, Eq.(\ref{33}) is not the final
definition of the averaging which thus appears somewhat more "subtle".

\subsection{The Strutinsky's averaging as a moving average\label{54}}

In Eq.(\ref{33}) the Fermi level $\lambda$ is supposed to be fixed, and the
polynomial $\overline{g}_{M\text{ , }\gamma}(\epsilon,\lambda)$ smoothes the
exact level density only in the vicinity of $\lambda$. Thus only a part of
the spectrum is " smoothed", i.e. the part $\epsilon\simeq\lambda$. To avoid
this drawback, it is necessary to consider $\lambda$ as a variable. The
averaged level density is thus defined as $\overline{g}_{M\text{ , }%
\gamma}(\epsilon,\lambda)$ making $\epsilon=\lambda$. This amounts to
perform a moving average (i.e. $\lambda$ is moved with $\epsilon$,
"centering" $\epsilon$ on $\lambda$).\newline
In the following, we shall call up the function $g_{_{M},\gamma}(\lambda,%
\lambda)$ as the Strutinsky's level density, and we will note it simply by $%
\overline{g}_{_{M},\gamma}(\lambda)$.\newline
Since the coefficients $c_{m}$ in (\ref{33}) depend on $\lambda$, the
Strutinsky's quantity $\overline{g}_{M\text{ , }\gamma}(\lambda)$ is, in
general, not a polynomial in $\lambda$. However, it is clear that, although
the Strutinsky level density is not really a polynomial, it behaves locally (%
$\sim\lambda\pm\gamma)$, like a least deviating polynomial approximation for
the exact level density $g_{o}(\lambda)$\ given by (\ref{1}).

Explicitly, we must replace $\epsilon$ with $\lambda$ in (\ref{33}):

\begin{gather}
\overline{g}_{M\text{ , }\gamma }(\lambda )=\dsum\limits_{n=0}^{\infty
}F_{M}\left( u_{n}\right) ,\text{ \ }u_{n}=\frac{\epsilon _{n}-\lambda }{%
\gamma }  \label{4} \\
F_{M}\left( x\right) =\widetilde{P}_{M}\left( x\right) \dfrac{1}{\gamma }%
\exp \left( -\text{ }x^{2}\right)  \notag \\
\text{\ }\widetilde{P}_{M}\left( x\right)
=\dsum\limits_{m=0}^{M}A_{m}H_{m}\left( x\right) ,\text{\ \ \ }A_{m}=\dfrac{%
H_{m}(0)}{m!2^{m}\sqrt{\pi }}  \notag \\
H_{m}(0)=(-1)^{m/2}.m!/\left( m/2\right) !\text{ \ if }m\text{ is even,} 
\notag \\
H_{m}(0)=0\text{ \ if }m\text{ is odd}  \notag
\end{gather}%
In (\ref{4}) the polynomial $\widetilde{P_{M}}$ constitutes the so-called
curvature correction term.

It is easy to check that the expression (\ref{4}) obtained from a
least-squares fitting, can be written as the usual folding procedure of the
exact level density \cite{13},\cite{17}:

\begin{equation}
\overline{g}_{_{M},\gamma}(\lambda)=\int\limits_{-\infty}^{\infty}g_{o}(%
\epsilon)F_{M}\left( \frac{\epsilon-\lambda}{\gamma}\right) d\epsilon
\label{400}
\end{equation}
which demonstrates the equivalence between the two points of view, i.e.
between the local least square smoothing (\ref{2}) and the averaging (\ref%
{400}).

Remark:\newline
It is to be noted that if $M$ increases to infinity and/or $\gamma$
decreases to zero, then $\overline{g}_{_{M},\gamma}$ tends toward $%
g_{o}(\epsilon)$. Obviously, in practice these parameters are finite.

\subsection{Necessary condition in the smoothing procedure\label{79}}

The least-square smoothing (\ref{2}) or its equivalent (\ref{400}) gives an
approximation \emph{to the exact level density }(\ref{1}). Therefore, if the
averaging is too accurate (see the previous remark) the procedure leads to a
curve which is very close to (\ref{1}). This curve remains characterized by
strong oscillations which express shell effects. However, the aim of the
Strutinsky method is precisely to remove these shell effects.

For $M=0$, the effective interval of averaging is governed by a pure
Gaussian (since $\widetilde{P}_{0}(x)=1$ in (\ref{4})). In order to wipe out
the shell effects from the averaging , the parameter $\gamma $ must be at
least of the order of the mean spacing between the shells.(denoted here by $%
\overline{\hbar \omega }$) near the Fermi level: 
\begin{equation}
\gamma \succsim \overline{\hbar \omega }  \label{7}
\end{equation}%
In this way, we obtain a "true" smooth density.

If Eq.(\ref{7}) is not satisfied, the level density remains characterized by
oscillations (quantum effects) and are in opposition with the character of
the semiclassical density.

For $M>0$, $F_{M}$ is not a pure Gaussian anymore, and, this case becomes
more complicated. Indeed, the width of the averaging function $F_{M}$ is not
really $\gamma $, because in (\ref{4}) the Gaussian is modulated by the
polynomial $\widetilde{P}_{M}.$\newline
It turns out that to maintain this width of the order of mean spacing shell
we have to enlarge the parameter $\gamma $ with respect to the order $M$.
Thus, the smoothing procedure implies a "coherency" between these two
parameters.

Usually, the couple ($\gamma,M$) is determined by the so-called plateau
condition \cite{17} from the typical ranges: $6\lesssim M\lesssim20,$ and, $%
\overline{\hbar\omega}\lesssim\gamma\lesssim2\overline{\hbar\omega}$.

Due to the above "coherency", it is easy to notice that the plateau is
invariably moved towards the right-hand side (toward the largest $\gamma )$,
when $M$ increases.

Since in the harmonic oscillator the spacing between the shells is constant,
we have in this simple case $\overline{\hbar \omega }=\hbar \omega $.

Fig.1 displays the Strutinsky level density (calculated from Rel.(\ref{4}))
for three values of the parameter $\gamma $.($M$ being fixed).

Curve( a): Since $\gamma $ is too small compared to $\hbar \omega $ the
curve (a) is characterized by strong oscillations (shell effects) which are
close to Dirac functions (see Rel.(\ref{1})).\newline
Curve (b): By increasing $\gamma $ one diminishes the magnitude of these
oscillations.\newline
Curve (c): In the third case, $\gamma $ is of the order of $\hbar \omega $,
the curve becomes smooth, and can be regarded as the mean behavior of the
exact level density (the so-called smooth component contained in the exact
level density (\ref{1})).

\subsection{Averaged particle-number, averaged energy, and shell correction}

The averaged particle-number and the averaged energy are defined through the
average density of states $\overline{g}_{M,\gamma}$ by:

\begin{gather}
\overline{N}_{M,\gamma }(\lambda )=\dint\limits_{-\infty }^{\lambda }%
\overline{g}_{M,\gamma }(\epsilon )d\epsilon  \label{10} \\
\overline{E}_{M,\gamma }(\lambda )=\dint\limits_{-\infty }^{\lambda
}\epsilon \overline{g}_{M,\gamma }(\epsilon )d\epsilon  \label{11}
\end{gather}%
The detailed expressions are given in Ref.\cite{8}.

In practice the upper bound of the integral giving the particle-number is
deduced from the equation

\begin{equation}
\overline{N}_{M,\gamma}(\lambda)=N_{0}  \label{498}
\end{equation}

where $N_{0}$ is the particle-number of the system. The quantity $\lambda$
is the Fermi level of the average density $\overline{g}_{M,\gamma}$ ( i.e.
the Strutinsky level density)

Finally the Strutinsky shell correction to the binding energy of the liquid
drop model is defined as follows :

\begin{equation}
\delta\overline{E}_{M,\gamma}=\dint \limits_{-\infty}^{\lambda_{0}}\epsilon
g_{o}(\epsilon)d\epsilon-\dint \limits_{-\infty}^{\lambda}\epsilon\overline{g%
}_{M,\gamma}(\epsilon)d\epsilon=\dsum \limits_{n=0}^{N_{0}}\epsilon_{n}-%
\overline{E}_{M,\gamma}(\lambda)  \label{48}
\end{equation}

Where $g_{o}(\epsilon)$ is the exact level density given by (\ref{1}) and $%
\lambda_{0}$ is its Fermi level (the last occupied level). Sometimes, $%
\overline{E}_{M,\gamma}(\lambda)$ is denoted as $\overline{\sum\epsilon_{n}}$%
.

In Rel.(\ref{48}), the shell correction should not contain any component of
the smooth energy which is by definition already included in the liquid drop
model. Consequently, the average density $\overline{g}_{M,\gamma}$ (or the
average energy $\overline{E}_{M,\gamma}(\lambda)$) must not contain any
residual shell effect.\newline
Moreover, since $\overline{g}_{M,\gamma}$is a least square approximation of $%
g_{o}$, one can write $g_{o}\approx g_{M,\gamma}+\delta g_{M,\gamma}$, with $%
\lambda_{0}\approx\lambda$. When the condition (\ref{7}) is fulfilled, $%
g_{M,\gamma}$ becomes smooth and the exact density $g_{o}$ oscillates around 
$g_{M,\gamma}$ making $\delta g_{M,\gamma}$ alternatively positive and
negative. Consequently, from (\ref{48}) one can deduce formally that:

\begin{equation}
\delta\overline{E}_{M,\gamma}(\lambda)\approx\dint
\limits_{-\infty}^{\lambda}\epsilon\delta g_{M,\gamma}(\epsilon)d\epsilon
\label{465}
\end{equation}
Thus, the oscillations of the shell correction $\delta\overline{E}_{M,\gamma
}(\lambda)$ as a function of $\lambda$ are due to the fluctuations of $%
\delta g_{M,\gamma}(\epsilon)$.

Note: \newline
In practice, due to the finite size of the spectrum, for the shell
correction, the following cut-off condition $\epsilon_{last\text{-}%
level}-\lambda\gg\gamma$ must be required (see \cite{17})

\section{The semiclassical level density}

\subsection{Bohr's correspondence principle}

Although the Strutinsky level density is mathematically well defined by (\ref%
{4}), there is no physical basis for this smoothing. Consequently, it is
necessary to build an other level density free of shell effects which would
be justified by physical arguments.

Shell effects are the consequence of the quantum nature of the level
distribution (\ref{1}). The natural way to eliminate such effects would be
to go over to the classical limit. To this end, we will apply the
correspondence principle (Bohr 1923) which states that the behavior of
quantum systems reduces to classical physics in "the limit of large quantum
numbers".\newline
Starting from this principle, the semiclassical level density can be deduced
by using the Euler-Maclaurin (EML) summation formula.\newline
In practice, this amounts to obtain an asymptotic series which contains only
three or four terms, all the others being divergent. The first term of this
series coincides with the known Thomas-Fermi approximation.\newline
The EML expansion used here is equivalent to the usual standard
semiclassical methods, e.g., the Wigner-Kirkwood expansion \cite{12},\cite%
{14},\cite{15}, or the method of the partition function \cite{19}. The
latter are based on asymptotic series of powers of $\hbar$. Indeed, in the
correspondence principle, the limit of large quantum numbers amounts to
taking the classical limit $\hbar\rightarrow0$.

\subsection{"Asymptotic limit of large quantum numbers"\label{687}}

In practical cases, the concept of "large quantum numbers" must be pr\'{e}%
cised by a more concrete definition. To illustrate this point, we start from
the typical example of the three-dimensional harmonic oscillator. Such
system is very simple, its energy levels are given by:

$\epsilon_{n}=(n+3/2)\hbar\omega$, $\ \ \ \ \ \ \ \ \ n$ $=0,1,2,3,..\infty$

The quantum number $n$ defines a shell, and $\hbar\omega$ represents the gap
between these shells.

Thus, $n$ is given by:

$n=\left( \epsilon_{n}-\epsilon_{0}\right) /\hbar\omega$ here $\epsilon
_{0}=\left( 3/2\right) \hbar\omega$ is the lowest level of the spectrum

The correspondence principle states that for large values of $n=\left(
\epsilon_{n}-\epsilon_{0}\right) /\hbar\omega$, the quantum physics reduces
to classical physics, in particular, the quantum (exact) level density $%
g_{o}(\lambda)$ defined by (\ref{1}) should approach the semiclassical level
density denoted here by $g_{sc}(\lambda)$. This can be written as :\newline
if $n=\left[ \left( \epsilon_{n}-\epsilon_{0}\right) /\hbar\omega\right]
\rightarrow\infty$, $\ \ \ \ $then $\ \ \ \ \ \ \ \ \ g_{o}\rightarrow
g_{sc} $

As already noted, the shell effects are mainly determined by the small part
of the spectrum $\left\{ \epsilon_{n}\right\} $ which is located in the
vicinity of the Fermi level $\lambda$. Consequently for these levels, we
have roughly $\epsilon_{n}\approx\lambda$, and, this limit becomes:\newline
if $n=\left[ \left( \lambda-\epsilon_{0}\right) /\hbar\omega\right]
\rightarrow\infty,\ \ \ \ $then$\ \ \ \ \ \ \ \ \ g_{o}(\lambda)\rightarrow
g_{sc}(\lambda)$

Since in general the Fermi level increases with the quantum numbers, the
arguments presented for the harmonic oscillator are also valid for any other
physical system. Therefore, we will consider the previous statement as
general. However, we must now redefine $\hbar \omega $\emph{\ }as the mean
shell-spacing in the neighborhood of the Fermi level\emph{\ }$\lambda $. As
in Rel.(\ref{7}), we denote it by $\overline{\hbar \omega }$.

\begin{equation}
\text{if \ \ \ \ \ }\frac{\left( \lambda-\epsilon_{0}\right) }{\overline{%
\hbar\omega}}\rightarrow\infty,\ \ \ \ \text{then}\ \ \ \ \ \ \ \ \
g_{o}(\lambda)\rightarrow g_{sc}(\lambda)  \label{4796}
\end{equation}

This limit is of course unphysical. Therefore in practice, we require the
following qualitative criterion:

\begin{equation}
\text{if \ $\frac{\lambda-\epsilon_{0}}{\overline{\hbar\omega}}\gg$1 \ \ \ \
\ then \ \ \ \ \ \ }\ g_{o}(\lambda)\approx g_{sc}(\lambda)  \label{55}
\end{equation}

We understand by (\ref{55}) that $\left[ \left( \lambda-\epsilon_{0}\right) /%
\overline{\hbar\omega}\right] $ is sufficiently large compared to the unity
so that $g_{o}(\lambda)$ can be considered as close as possible to $%
g_{sc}(\lambda)$ with a "satisfactory accuracy".

Finally, the "asymptotic limit of large quantum numbers" can be defined
theoretically by (\ref{4796}), or practically by (\ref{55}).

Notes:\newline
In practical cases, $\lambda\gg\epsilon_{0}$, therefore the previous
requirement can be replaced with $\lambda/\overline{\hbar\omega}%
\gg\nolinebreak1$. In this case, the Fermi level $\lambda$ must be measured
from the bottom of the well

\subsection{Two well-known analytical cases}

The procedure described above is applied in appendix B to two simple cases.

\subsubsection{The semiclassical level density of the harmonic oscillator}

The eigenenergies of the isotropic oscillator are :\newline
\begin{gather}
E_{n_{x},n_{y},n_{z}}=(n_{x}+n_{y}+n_{z}+\frac{3}{2})\hbar \omega _{0}
\label{78} \\
n_{x},n_{y},n_{z}=0,1,2,.....\infty  \notag
\end{gather}

The semiclassical level density of the harmonic oscillator is a simple
parabola (see appendix B):

\begin{equation}
g_{sc.}(\lambda)=\frac{1}{2}\frac{\lambda^{2}}{\left( \hbar\omega_{0}\right)
^{3}}-\frac{1}{8}\frac{1}{\hbar\omega_{0}}  \label{16}
\end{equation}

This result is well-known, and was established very early by means of the
partition function \cite{19}, or more recently by the Wigner-Kirkwood
expansion \ \cite{20}.

\subsubsection{The semiclassical level density of the infinitely deep cubic
box}

For the case of cubic box with totally reflecting walls the spectrum is
given by:

\begin{gather}
E_{n_{x}n_{y}n_{z}}=\left( n_{x}^{2}+n_{y}^{2}+n_{z}^{2}\right) E_{0}
\label{21} \\
E_{0}=\pi ^{2}\hbar ^{2}/\left( 2ma_{0}^{2}\right) ,\text{ }a_{0}\text{=
side of the cubic box}  \notag \\
n_{x},n_{y},n_{z}=1,2,.....\infty \text{\ }  \notag
\end{gather}%
The semiclassical level density of the infinite cubic box (see appendix B)
is also an "old " result \cite{19}, \cite{21}.

\begin{equation}
g_{sc.}(\lambda)\approx\frac{1}{E_{0}^{3/2}}\frac{\pi}{4}\sqrt{\lambda}-%
\frac{3\pi}{8}\frac{1}{E_{0}}+\frac{3}{8}\frac{1}{E_{0}^{1/2}}\frac{1}{\sqrt{%
\lambda}}  \label{18}
\end{equation}

\subsection{Semiclassical shell correction}

We define the semiclassical energy by a very similar relations to (\ref{10}-%
\ref{48}), replacing the Strutinsky level density by the one of the
semiclassical density. The corresponding Fermi level is denoted as $\lambda
_{sc}$.%
\begin{gather}
N_{sc}(\lambda _{sc})=\dint\limits_{-\infty }^{\lambda _{sc}}g_{sc}(\epsilon
)d\epsilon  \label{19} \\
E_{sc}(\lambda _{sc})=\dint\limits_{-\infty }^{\lambda _{sc}}\epsilon
g_{sc}(\epsilon )d\epsilon  \label{25}
\end{gather}

\begin{equation}
N_{sc}(\lambda_{sc})=N_{0}
\end{equation}

\begin{equation}
\delta E_{sc}=\dint \limits_{-\infty}^{\lambda_{0}}\epsilon
g_{o}(\epsilon)d\epsilon-\dint \limits_{-\infty}^{\lambda_{sc}}\epsilon
g_{sc}(\epsilon)d\epsilon=\dsum
\limits_{n=0}^{N_{0}}\epsilon_{n}-E_{sc}(\lambda_{sc})  \label{95}
\end{equation}

One must note that unlike $\delta\overline{E}_{M,\gamma}(\lambda)$, the
quantity $\delta E_{sc}$ does not depend on any free parameter.

\section{The connection between the "Strutinsky's level density" and the
semi-classical level density.}

\subsection{Assumptions and quantitative approach of the asymptotic limit 
\label{468}}

We know from the Bohr principle given in subsect. \ref{687} that in the
"asymptotic limit" (\ref{55}) we must have the approximation $g_{o}(\lambda
)\simeq g_{sc}(\lambda )$. Since the Strutinsky averaging (\ref{400}) gives
an approximation to the exact level density $g_{o}(\lambda )$, normally in
this limit it should also give an approximation to the semiclassical level
density $g_{sc}(\lambda )$. The role of the curvature correction term would
be to improve the approximation.

We start from the averaging (\ref{400}), substituting $g_{sc}(\lambda)$ for $%
g_{o}(\lambda)$ in the "asymptotic limit"(\ref{55}). In the Strutinsky
averaging the parameter $\gamma$ must be of the order of the mean shell
spacing $\overline{\hbar\omega}$ near the Fermi level (see Eq.(\ref{7})).
Consequently, in Rel.(\ref{55})\ we should replace $\overline{\hbar\omega}$
with $\gamma$.

\begin{gather}
\overline{g}_{_{M},\gamma }(\lambda )\approx \dint\limits_{-\infty }^{\infty
}g_{sc}(\epsilon )\widetilde{P_{M}}\left( \dfrac{\epsilon -\lambda }{\gamma }%
\right) \dfrac{1}{\gamma }\exp \left( -\text{ }\left( \dfrac{\epsilon
-\lambda }{\gamma }\right) ^{2}\right) d\epsilon  \label{4444} \\
\text{with }\frac{\lambda -\epsilon _{0}}{\gamma }\gg 1\text{, \ \ and }%
\gamma \succsim \overline{\hbar \omega }
\end{gather}

Making $X=\dfrac{\epsilon-\lambda}{\gamma}$, we obtain:

$\overline{g}_{_{M},\gamma}(\lambda)\approx\dint
\limits_{-\infty}^{\infty}g_{sc}(\lambda+\gamma X)\widetilde{P_{M}}\left(
X\right) \exp\left( -X^{2}\right) dX,$ \ \ \ \ $M$ even

Now, one replaces the semiclassical density $g_{sc}(\lambda+\gamma X)$ by
its $\left( M+2\right) $ first terms of the Taylor expansion around $\lambda 
$\ ($M$ must be even in $\widetilde{P}_{M}\left( X\right) $). The last term
gives an estimation of the remainder:

\begin{multline*}
\overline{g}_{_{M},\gamma }(\lambda )\approx \int\limits_{-\infty }^{\infty
}[g_{sc}(\lambda )+\dsum\limits_{k=1}^{M+1}\dfrac{\left( \gamma X\right) ^{k}%
}{k!}\dfrac{d^{k}g_{sc}(\lambda )}{d\lambda ^{k}} \\
+\dfrac{\left( \gamma X\right) ^{M+2}}{\left( M+2\right) !}\dfrac{%
d^{M+2}g_{sc}(\lambda )}{d\lambda ^{M+2}}]\widetilde{P}_{M}\left( X\right)
\exp \left( -X^{2}\right) dX
\end{multline*}

It is easy to show that $\widetilde{P}_{M}\left( X\right) \exp \left(
-X^{2}\right) $ behaves like a delta function with respect to any polynomial
of order $k\leq M$. Consequently the first term gives back $g_{sc}(\lambda )$%
, and the second has no contribution (by noticing that $X^{M+1}$ is odd).
The remaining integral, i.e.:

$I(M)=\int\limits_{-\infty}^{\infty}X^{M+2}\widetilde{P}_{M}\left( X\right)
\exp\left( -X^{2}\right) dX$

is obtained from appendix A.

Finally:

\ 
\begin{gather}
\overline{g}_{_{M},\gamma }(\lambda )\approx g_{sc}(\lambda )\left\{
1+R_{_{M},\gamma }(\lambda )\right\}  \label{49} \\
R_{_{M},\gamma }(\lambda )=\gamma ^{M+2}\dfrac{C_{M+2}}{\left( M+2\right) !}%
\frac{1}{g_{sc}(\lambda )}\dfrac{d^{M+2}g_{sc}(\lambda )}{d\lambda ^{M+2}} 
\notag \\
C_{M+2}=(-1)^{M/2}\frac{1.3.5.....(M+1)}{2^{(M+2)/2}}\text{ , \ \ }M\text{
even,}  \notag \\
\text{\textit{with} \ \ \ }\lambda -\epsilon _{0}\gg \gamma \succsim 
\overline{\hbar \omega }  \notag
\end{gather}

Eq.(\ref{49}) is fundamental and gives the straightforward link between the
semiclassical level density $g_{sc}(\lambda)$, and the Strutinsky level
density $\overline{g}_{_{M},\gamma}(\lambda)$ in the asymptotic limit $%
\lambda-\epsilon_{0}\gg\gamma$, with the necessary condition of the
smoothing procedure $\gamma\succsim\overline{\hbar\omega}$. \newline
It should be noted that it is $\overline{g}_{_{M},\gamma}(\lambda)$ which is
deduced from $g_{sc}(\lambda)$ and not the opposite. Moreover, $%
g_{sc}(\lambda)$ does not depend on any free parameter. Therefore $%
g_{sc}(\lambda)$ must be considered as the "true" smooth level density (the
so-called smooth component of the quantum density (\ref{1})), and $\lambda$
is its Fermi level.\newline
The quantity $R_{_{M},\gamma}(\lambda)$ is the remainder of the averaging.
From (\ref{49}), it can easily be identified with the relative error:

\begin{equation*}
\left\vert R_{_{M},\gamma}(\lambda)\right\vert \approx\left\vert \dfrac {%
g_{sc}(\lambda)-\overline{g}_{_{M},\gamma}(\lambda)}{g_{sc}(\lambda )}%
\right\vert
\end{equation*}
\newline
and represents the noise (for the density of states) of the Strutinsky
method. In actual calculations, it is implicit, and thus unknown. It is
contained intrinsically in $\overline{g}_{_{M},\gamma}(\lambda)$.

It is easy to check from (\ref{49}) that the coefficient $C_{M+2}/\left(
M+2\right) !$ in the remainder decreases theoretically to zero as $M$
increases to infinity ($\lambda $,$\gamma $ being fixed), provided that $%
g_{sc}(\lambda )$ is sufficiently regular. This in principle improves the
average. Nevertheless, we have seen in subsec.\ref{79} that large values of $%
M$ involve necessary a slight increasing of $\gamma $ in the smoothing
procedure, which in turn increases somewhat the remainder as it can easily
be seen from (\ref{49}). Thus, it is not possible to reduce the remainder
without ending.

In practical cases, the "optimal choice" $M\sim16-30$ leads to a very good
precision, i.e. $\left\vert R_{_{M},\gamma}(\lambda)\right\vert <0.01$.

From Rel.(\ref{49}) it is clear that the dependency on the two parameter $%
(M,\gamma )$ becomes more and more weak as the remainder decreases to zero.%
\newline
The only one special case where the remainder vanishes rigorously is that
one where $g_{sc}(\lambda )$ is a pure polynomial of degree less or equal to 
$M$. This happens in the harmonic oscillator case. For this reason, the
Strutinsky method must not be "judged" in this example when $M\geqq 2$.

Fundamentally, the Strutinsky level density appears in (\ref{49}) only as an
approximation (and thus is not strictly equivalent as it is often claimed)
to the semiclassical level density. Consequently, the smooth Strutinsky
energy Eq.(\ref{11}), and the Strutinsky shell correction of Eq.(\ref{48})
must also be considered as an approximations to the respective semiclassical
quantities given by Eq.(\ref{25}) and Eq.(\ref{95}).

In fact, we shall see in the next subsection that the remainder tends also
to zero with $\lambda$ like $\left( \gamma/\lambda\right) ^{M+2}$, and thus
becomes negligible only in the "asymptotic limit" $\left( \gamma
/\lambda\right) \ll1$.

\subsection{The relative error on the Strutinsky level density in two
special cases \label{580}}

It is instructive, to apply our result (\ref{49}) for the cases seen
previously, i.e., the harmonic oscillator (\ref{16}) and the cubic box (\ref%
{18}) with the previous assumptions $\lambda-\epsilon_{0}\gg \gamma\succsim%
\overline{\hbar\omega}$. For these calculations we choose $M=0$ and $M=2$ ($%
M $ must be even).\newline
For the harmonic oscillator case, using \ $g_{sc}(\lambda)$ from Eq.(\ref{16}%
), we find:

For $M=0$

\begin{equation}
\overline{g}_{_{M=0},\gamma}(\lambda)\approx g_{sc}(\lambda)\left[ 1+\dfrac{1%
}{2}\dfrac{\gamma^{2}}{\lambda^{2}}\right] \text{, \ \ \textit{with} \ \ \ }%
\lambda\gg\gamma\succsim\overline{\hbar\omega}
\end{equation}
(Here of course, the mean spacing between the shells is constant, and we
have simply $\overline{\hbar\omega}=\hbar\omega$).

For $M=2$

\begin{equation}
\overline{g}_{_{M=2},\gamma}(\lambda)\approx g_{sc}(\lambda)\left[ 1+0\right]
\text{, \ \ \textit{with} \ \ \ }\lambda\gg\gamma\succsim \overline{%
\hbar\omega}
\end{equation}

Since the semiclassical level density is a parabola, the derivative which
appears in the remainder $R_{_{M},\gamma}$ in Eq.(\ref{49}) cancels for $%
M\geq2$, therefore one obtains the exact result (the remainder is $0$). One
should not be too "impressed" by this case ( see subsection \ref{468}).

For the infinite cubic box , using $g_{sc}(\lambda)$ from Eq.(\ref{18}) we
get for $M=0$ and $M=2$:

\begin{equation}
\overline{g}_{_{M=0},\gamma}(\lambda)\approx g_{sc}(\lambda)\left[ 1-\dfrac{1%
}{16}\dfrac{\gamma^{2}}{\lambda^{2}}\right] \text{, \ \ \textit{with} \ \ \ }%
\lambda\gg\gamma\succsim\overline{\hbar\omega}  \label{473}
\end{equation}

\begin{equation}
\overline{g}_{_{M=2},\gamma}(\lambda)\approx g_{sc}(\lambda)\left[ 1+\dfrac{%
15}{512}\frac{\gamma^{4}}{\lambda^{4}}\right] \text{, \ \ \textit{with} \ \
\ }\lambda\gg\gamma\succsim\overline{\hbar\omega}  \label{474}
\end{equation}

There also, for both cases, we obtain very similar relations.

Thus, in these four cases, the Strutinsky level density approaches the
semiclassical level density, and the relative error (remainder) tends to
zero only in the asymptotic limit $\gamma/\lambda\ll\nolinebreak1$. Moreover
it is clear that in this limit, the Strutinsky densities becomes practically
independent on the smoothing parameters $\left( \gamma,M\right) $.

In realistic cases, the Fermi level $\lambda$ is fixed for a given nucleus.
It turns out that for medium and heavy nucle\emph{i}, the quantity $\lambda$
lies several units of $\overline{\hbar\omega}$ above the bottom of the well
, since $\overline{\hbar\omega}\approx\gamma$, the quotient $\gamma/\lambda$
is thus small and the remainder is practically negligible. Consequently, for
these cases, the Strutinsky density of states is very close to the
semiclassical level density.

The relative error of the Strutinsky density for the cubic box is
illustrated in Fig. 2.

In Fig.2a we compare the "numerical" Strutinsky level density (curve a)
calculated by means of Eq.(\ref{4}) to the semiclassical density (curve b)
given by Eq.(\ref{18}). Apart from the region near zero (very small $\lambda 
$), and in spite of the "low" order $M=0$ of the curvature correction, It is
clear, that the two densities are practically indistinguishable. The
theoretical link between both densities is given by Eq.(\ref{473})\newline
In Fig.2b, we can see that the "numerical" relative error $(g_{sc}-\overline{%
g}_{_{M=0},\gamma })/g_{sc}$ of the Strutinsky density (denoted by (b-a)/b)
with respect to the semiclassical density (regarded as the true smooth
density), is very small, especially in the asymptotic limit ($\left( \gamma
/\lambda \right) \ll 1$). In the latter, this error becomes close to the
theoretical value $\gamma ^{2}/16\lambda ^{2}$ given by Eq.(\ref{473}).

\subsection{The relative error on the Strutinsky energy in the two previous
cases\label{796}}

The average energy $\overline{E}_{M,\text{ }\gamma}(\lambda)$ can be deduced
by combining Eq.(\ref{11}) and Eq.(\ref{49}) with the assumptions of the
asymptotic limit and the necessary condition of smoothing made in Eq.(\ref%
{49}):

\begin{gather}
\overline{E}_{M,\text{ }\gamma }(\lambda )\approx E_{sc}\left( \lambda
\right) \left[ 1+\rho _{_{M},\gamma }\left( \lambda \right) \right]
\label{7411} \\
\rho _{_{M},\gamma }\left( \lambda \right) =\frac{S_{_{M},\gamma }\left(
\lambda \right) }{E_{sc}(\lambda )} \\
S_{_{M},\gamma }\left( \lambda \right) =\dfrac{\gamma ^{M+2}C_{M+2}}{\left(
M+2\right) !}\int\limits_{-\infty }^{\lambda }\epsilon \dfrac{%
d^{M+2}g_{sc}(\epsilon )}{d\epsilon ^{M+2}}d\epsilon  \label{7900} \\
\text{\textit{with} \ \ \ }\lambda -\epsilon _{0}\gg \gamma \succsim 
\overline{\hbar \omega }
\end{gather}

where $\rho_{_{M},\gamma}\left( \lambda\right) $ and $S_{_{M},\gamma}\left(
\lambda\right) $ are respectively the relative and the absolute errors on
the smooth (Strutinsky) energy $\overline{E}_{M,\text{ }\gamma}(\lambda)$.

We know from Eq.(\ref{49}) that $\overline{E}_{M,\text{ }\gamma }(\lambda )$
must be considered as an approximation to $E_{sc}(\lambda )$. Besides,
unlike $\overline{E}_{M,\text{ }\gamma }(\lambda )$, the quantity $%
E_{sc}(\lambda )$ does not depend on any unphysical parameter. As before the
remainder $\rho _{_{M},\gamma }\left( \lambda \right) $ of the Strutinsky
energy must be related to the relative error:

\begin{equation}
\left\vert \rho_{_{M},\gamma}\left( \lambda\right) \right\vert
\approx\left\vert \dfrac{E_{sc}\left( \lambda\right) -\overline{E}_{M,\text{ 
}\gamma}(\lambda)}{E_{sc}\left( \lambda\right) }\right\vert  \label{11111}
\end{equation}

Once again, in order to illustrate some features of the Strutinsky method we
apply the above result for the harmonic oscillator and for the cubic box
with $M=0$ and $M=2$.

For the harmonic oscillator:

\begin{equation}
\overline{E}_{M=0,\text{ }\gamma}(\lambda)\approx E_{sc}(\lambda)\left[ 1+%
\dfrac{\gamma^{2}}{\lambda^{2}}\right] \text{, \ \ \textit{with} \ \ \ }%
\lambda\gg\gamma\succsim\overline{\hbar\omega}  \label{5910}
\end{equation}

\begin{equation}
\overline{E}_{M=2,\text{ }\gamma }(\lambda )\approx E_{sc}(\lambda )\left[
1+0\right] \text{, \ \ \textit{with} \ \ \ }\lambda \gg \gamma \succsim 
\overline{\hbar \omega }
\end{equation}%
Where $E_{sc}(\lambda )$ of the harmonic oscillator is given in appendix B.%
\newline
In the same way, we obtain for the cubic box:

\begin{equation}
\overline{E}_{M=0,\text{ }\gamma}(\lambda)\approx E_{sc}(\lambda)\left[ 1-%
\frac{5}{16}\dfrac{\gamma^{2}}{\lambda^{2}}\right] \text{, \ \ \textit{with}
\ \ \ }\lambda\gg\gamma\succsim\overline{\hbar\omega}  \label{9173}
\end{equation}

\begin{equation}
\overline{E}_{M=2,\text{ }\gamma }(\lambda )\approx E_{sc}(\lambda )\left[ 1-%
\frac{25}{512}\dfrac{\gamma ^{4}}{\lambda ^{4}}\right] \text{, \ \ \textit{%
with} \ \ \ }\lambda \gg \gamma \succsim \overline{\hbar \omega }
\end{equation}%
Where $E_{sc}(\lambda )$ of the cubic box is also given in appendix B.

Thus, in the asymptotic limit $\left( \gamma/\lambda\right) \ll1$ (i.e. for
medium and \ heavy nuclei), as for the density of states (\ref{49}), the
relative error is small, and we have also $\overline{E}_{M,\text{ }\gamma
}(\lambda)\approx E_{sc}(\lambda)$. Thus, for the smooth energy, the
Strutinsky method is a good approximation of the semiclassical method. A
straightforward consequence is that the smooth (Strutinsky) energy $%
\overline{E}_{M,\text{ }\gamma}$ becomes practically independent on the
smoothing parameters $\left( M,\gamma\right) $ in this limit.

We give in Fig. 3 an illustration of the relative error for the Strutinsky
energy for the cubic box

Fig.3a: Same conclusions as Fig.2a, i.e. $\overline{E}_{M=0,\text{ }\gamma
}(\lambda )$ and $E_{sc}(\lambda )$ are indistinguishable\newline
Fig.3b: There also, the relative error on the Strutinsky energy (with
respect to the semiclassical energy) tends toward zero in the asymptotic
limit $\left( \gamma /\lambda \right) \ll 1$\nolinebreak\ and approaches the
theoretical value $5\gamma ^{2}/16\lambda ^{2}$ given by Eq.(\ref{9173}).

\subsection{The new understanding of the plateau condition on the average
(Strutinsky) energy}

As seen before, the relative error plays a major role in the Strutinsky
energy.

From Rel.(\ref{11111}), it is clear that if $\left\vert \rho_{_{M},\gamma
}\left( \lambda\right) \right\vert \ll1$, the relative variations of the
Strutinsky energy $\overline{E}_{M,\text{ }\gamma}(\lambda)$ \emph{are small
compared to} $\overline{E}_{M,\text{ }\gamma}(\lambda)$ itself (or to $%
E_{sc}(\lambda)$).

For instance, if one plots $\overline{E}_{M,\text{ }\gamma}(\lambda)$ as a
function of the parameter $\gamma$ ($M$ and $\lambda$ being constant) it
appears a "plateau" in the graph. This means that $\overline{E}_{M,\text{ }%
\gamma}(\lambda)$ is "practically constant at the scale of its own value"
(at least in the interval $\ \lambda\gg\gamma\succsim\overline{\hbar\omega}$%
), i.e.:

\begin{equation}
\left( \frac{\Delta\overline{E}_{M,\text{ }\gamma}(\lambda)}{\overline {E}%
_{M,\text{ }\gamma}(\lambda)}\right) _{\lambda\gg\gamma\succsim \overline{%
\hbar\omega}}\ll1  \label{22222}
\end{equation}
(which is close to (\ref{11111})). This does not necessarily mean that the
derivative of $\overline{E}_{M,\text{ }\gamma}(\lambda)$ cancels as in the
traditional plateau condition \cite{17}. The same remark holds for the
Strutinsky density.

We must point out that the relative error is proportional to the quantity $%
\left( \gamma/\lambda\right) ^{M+2}$, so that the plateau is improved at
large values of $M.$

Fig. 4 shows an illustration of the plateau (defined by (\ref{22222})) for
the energy in the cubic box case

Fig.4a: The Strutinsky energy $\overline{E}_{M,\text{ }\gamma }$ of the
cubic box is plotted as function of $\gamma $ for four values of the order $%
M $. The particle-number is fixed arbitrarily at $N_{0}=200$ with a Fermi
level $\lambda =64.255E_{0}$. It is clear that the fluctuations $\Delta 
\overline{E}_{M,\text{ }\gamma }$ are small compared to $\overline{E}_{M,%
\text{ }\gamma }$. At this "scale" a clear plateau is noticed.\newline
Fig.4-b: The same figure as Fig.4-a at a "reduced scale" shows the important
variations of the plateau, especially on the r.h.s of the figure. If we
continue to "zoom in" the curve, it appears several minima and maxima (i.e. $%
\partial \overline{E}_{M,\text{ }\gamma }/\partial \gamma =0$). Some of
them, have nothing to do with a plateau. For example, in the vicinity of $%
\gamma \approx 154$ $E_{0}$ a minimum occurs for the order $M=16$ which does
not really belong to any plateau. Thus our "macroscopic" definition of the
plateau seems more "adapted" than the old version based on the
"stationarity" of $\overline{E}_{M,\text{ }\gamma }$ with respect to $\gamma 
$. In fact, it contains implicitly the concept of the relative error which
plays a central role in the numerical applications in the method. Indeed the
minimization of the relative error (see below) avoids the ambiguity of the
(old) plateau condition, because a stationary point is not necessary a
plateau, whereas the minimization of the relative error leads indisputably
to the true value (at the very least to the optimal value) of the smooth
energy, an thence of the shell correction (see below).

\subsection{The Strutinsky shell correction\label{4976}}

\subsubsection{The critical point of the Strutinsky method}

First, one must recall that the Strutinsky shell correction and the
semiclassical shell correction are respectively defined through Rel.(\ref{48}%
) and (\ref{95}), i.e.

$\delta \overline{E}_{M,\gamma }=\sum\limits_{occupied}\epsilon _{n}-%
\overline{E}_{M,\gamma }(\lambda )$,

$\delta E_{sc}=\sum\limits_{occupied}\epsilon _{n}-E_{sc}(\lambda )$

Subtracting the second equation from the first, and using Rel.(\ref{7411}),
one obtains a straightforward relation between these two quantities.:

\begin{equation}
\delta\overline{E}_{M,\text{ }\gamma}(\lambda)\approx\delta
E_{sc}(\lambda)-S_{_{M},\gamma}\left( \lambda\right)  \label{7413}
\end{equation}

Since $\overline{E}_{M,\gamma }(\lambda )$ is considered as an approximation
to $E_{sc}(\lambda )$ (see previous subsection), the Strutinsky shell
correction $\delta \overline{E}_{M,\text{ }\gamma }(\lambda )$ must be also
regarded as an approximation to the semiclassical shell correction $\delta
E_{sc}(\lambda )$. In this respect, $S_{_{M},\gamma }\left( \lambda \right) $
represents the same absolute error in both formulae (\ref{7411},\ref{7413}).
However, the essential point is that, this error has not the same importance
in these two results \newline
Indeed, the two shell corrections $\delta E_{M,\text{ }\gamma }^{{}}(\lambda
)$, and $\delta E_{sc}(\lambda )$ are obtained as the difference between two
close large numbers (i.e. the sum of single-particle energies and their
averages) therefore they are significantly smaller compared to these
quantities (i.e. $\overline{E}_{M,\text{ }\gamma }$ or $E_{sc}$).\newline
For example, the Ref \cite{11} gives for the case of $^{154}$Sn (neutrons)
the typical "realistic" values $\tsum\nolimits_{{}}\epsilon _{i}=-1122.5$ $%
MeV$, $\overline{E}_{M,\text{ }\gamma }=-1132.0$ $MeV$ (the order $M$ is not
pr\'{e}cised in that work), and hence, the shell correction $\delta 
\overline{E}_{M,\text{ }\gamma }=9.5$ $MeV$ is thus, much smaller than $%
\overline{E}_{M,\text{ }\gamma }$. Consequently, the same absolute error $%
S_{_{M},\gamma }\left( \lambda \right) $ which is relatively small in Eq.(%
\ref{7411}) might become non negligible in Eq.(\ref{7413}) for the
Strutinsky shell correction.\newline
In addition, in a number of cases the shell correction might also become so
small that the relative error (\emph{in the shell correction}) has no longer
a sense. Thus, for the shell correction, the relative error does not play
the same leading role as for the Strutinsky energy (or the Strutinsky
density), so that the (Strutinsky) shell correction might become strongly
dependent to the choice of the parameter $\gamma $. This means that the
error could exceed the shell correction itself if this error is not
optimized (i.e. minimized).

\subsubsection{The optimization of the method with the "rule of the relative
remainder R.R.R".}

The shell correction is defined as the difference between two quantities $%
\tsum\nolimits_{{}}\epsilon _{i}$, and $\overline{E}_{M,\text{ }\gamma
}(\lambda )$. Only the latter depends on the parameter $\gamma $ (and also $M
$) through the remainder $\rho _{_{M},\gamma }\left( \lambda \right) $ from (%
\ref{7411}). By minimizing this remainder (i.e. the relative error) with
respect to $\gamma $, we make $\overline{E}_{M,\text{ }\gamma }(\lambda )$
as close as possible to $E_{sc}$, therefore we make $\delta \overline{E}_{M,%
\text{ }\gamma }(\lambda )$ as close as possible to $\delta E_{sc}(\lambda )$
(i.e. the true shell correction). Thus, the minimization of the relative
error made on $\overline{E}_{M,\text{ }\gamma }(\lambda )$, should lead to
the independency ( or at least to a weak dependence) of $\delta \overline{E}%
_{M,\text{ }\gamma }(\lambda )$ on the parameters $\left( \gamma ,M\right) $%
. Thence, we can affirm that it is the minimization of this relative error
which is the source of the plateau, not the opposite. This should be the
most appropriate way for finding the true (or the best) value of the shell
correction. Fig.5 displays a practical illustration of this minimization
(the so-called "rule of relative remainder").

We show in Fig.5a how to find the optimal value for the parameter $\gamma$
in the cubic box case. We again consider in Fig.5a the case given in Fig.4a%
\newline
The steps are the following:

\begin{itemize}
\item First we calculate the Strutinsky energy at the Fermi level $\overline{%
E}_{M,\text{ }\gamma}(\lambda)$ as a function of the parameter $\gamma$ ($%
\lambda$ being fixed), for some close values of the order $M$.

\item We must minimize the remainder of Rel.(\ref{11111}) as follows:
\end{itemize}

\begin{eqnarray*}
\left\vert \frac{\partial }{\partial \gamma }\rho _{_{M},\gamma }\left(
\lambda \right) \right\vert &=&\left\vert \frac{\partial }{\partial \gamma }%
\dfrac{E_{sc}\left( \lambda \right) -\overline{E}_{M,\text{ }\gamma
}(\lambda )}{E_{sc}\left( \lambda \right) }\right\vert \approx 0 \\
&=&\left\vert \dfrac{\dfrac{\partial \overline{E}_{M,\text{ }\gamma
}(\lambda )}{\partial \gamma }}{E_{sc}\left( \lambda \right) }\right\vert
\approx \left\vert \dfrac{\dfrac{\partial \overline{E}_{M,\text{ }\gamma
}(\lambda )}{\partial \gamma }}{\overline{E}_{M,\text{ }\gamma }(\lambda )}%
\right\vert
\end{eqnarray*}

\begin{itemize}
\item Then, this quantity is plotted as a function of $\gamma $ for each
value of $M$. Thence, we should look for the minimum of this function
(relative error).
\end{itemize}

However for a fixed $M$, this function has an oscillatory behavior (around
zero) which leads to several local minima. Nevertheless, due to the fact
that these curves do not cancel simultaneously, it is possible to remove
these unpleasant oscillations by considering the mean relative error over
some (relative close) values of $M$.\newline
The mean relative error (over $M=16,20,24,28$) on the Strutinsky energy is
plotted against $\gamma$. The minimum (optimal) value is found to be about $%
\gamma_{opt}\approx28.6E_{0}$ and corresponds effectively to the best value
(see Fig.5b)\newline
In Fig.5b, we note a good agreement between the true value (semiclassical)
given by the upper straight line $\overline{E}_{sc}\approx8093.97E_{0}$ and
the Strutinsky values. In fact, for the optimized $\gamma$, the calculated
values are $\overline{E}_{16,\text{ }\gamma_{opt}}\approx8094.52E_{0}$,$%
\overline{\text{ }E}_{20,\text{ }\gamma_{opt}}\approx8094.64E_{0}$, $%
\overline{E}_{24,\text{ }\gamma_{opt}}\approx8095.21E_{0}$, $\overline{E}%
_{28,\text{ }\gamma_{opt}}\approx8096.31E_{0}$, for $M=16,20,24,28.$\newline
At last the sum of single-particle energies is $\tsum
\nolimits_{occ.states}\epsilon_{i}\approx7842E_{0}$.

The true shell correction $\delta\overline{E}_{sc}\approx-251.97E_{0}$, and
the corresponding Strutinsky shell corrections are thus: $\delta\overline {E}%
_{16,\text{ }\gamma_{opt}}\approx252.52E_{0}$, $\delta\overline{\text{ }E}%
_{20,\text{ }\gamma_{opt}}\approx252.64E_{0}$, $\delta\overline {E}_{24,%
\text{ }\gamma_{opt}}\approx253.21E_{0}$, $\delta\overline {E}_{28,\text{ }%
\gamma_{opt}}\approx254.31E_{0}$, which are in good agreement with the true
(exact) value. Without optimization the results of the Strutinsky shell
correction will certainly be random.

\subsection{Case of realistic wells:}

\subsubsection{The Strutinsky level density}

First, we must note that the spectrum of the finite potential is composed of
discrete negative levels plus a continuum. For this potential, the
definition of the exact level density (\ref{1}) must be modified by adding
an appropriate continuous expression $g_{c}(\epsilon )$:

\begin{equation}
g_{o}(\epsilon)=\dsum
\limits_{n}^{{}}\delta(\epsilon-\epsilon_{n})+g_{c}(\epsilon)  \label{27}
\end{equation}

For spherical potentials, the continuum is defined by the scattering phase
shift, whereas for the deformed case it can be solved by the more
complicated S-matrix method (see \cite{7}).

Next, one recalls that the result (\ref{49}) (which is the basis of the
present work) comes from Eq.(\ref{400}). The latter is valid for any smooth
potential regardless whether it is infinite or not. Indeed, it is to be
noted that the interval of averaging in (\ref{400}) goes from $-\infty$ up
to $+\infty$ so that the preceding demonstration remains still valid for a
finite well. One has simply to add the continuum of (\ref{27}) to the
discrete spectrum into this integral. Thence, as for infinite potentials, it
is clear that the Strutinsky's level density should also be an approximation
to the semiclassical level density for finite wells.

In practice, the rigorous treatment of the continuum is not easy. The
standard recipe consists of using the discrete positive energies to
"simulate" this continuum \cite{8}. These energies are usually obtained by
diagonalizing the Hamiltonian matrix in a truncated harmonic oscillator
basis. In fact, this delicate problem seems now to be solved by the
so-called GFOE method (Green's function oscillator expansion) \cite{11}
which improves the standard method..

\subsubsection{The semiclassical level density}

It is well-known that the level density of finite potential becomes singular
at the top of the well \cite{23},\cite{7}. For this reason, it is not
possible to find a local polynomial approximation to the semiclassical level
density in the vicinity of this singularity. In other words, the least
square averaging (\ref{2}) does not hold in this case, precisely because $%
g_{sc}(\lambda )\rightarrow\infty$ , as $\lambda\rightarrow0$, i.e. at the
zero-energy. This explains why the two methods do not lead to the same
results, especially for the weakly bound nuclei. Far from the zero-energy
there is no problem.

\subsubsection{Comparison between the two level densities}

It would be interesting to determine the limit where the Strutinsky level
density deviates significantly from the semiclassical density. To this end,
we will be comparing numerically the Strutinsky level density to the
semiclassical ( Wigner-Kirkwood) level density by employing the\ result of
Ref \cite{22}:

\begin{equation}
g_{sc}(\lambda)=\dfrac{dN_{sc}(\lambda)}{d\lambda_{sc}}  \label{951}
\end{equation}

with:

\begin{gather}
N_{sc}=\frac{1}{3\pi ^{2}}\left( \frac{2m}{\hbar ^{2}}\right)
^{3/2}\dint\limits_{0}^{r_{sc}}d^{3}r\left[ (\lambda _{sc}-U)^{3/2}+\left( 
\frac{2m}{\hbar ^{2}}\right) ^{-1}\Omega \right]  \label{798} \\
\Omega =\left[ \frac{3}{4}\kappa ^{2}\left( \overset{\rightarrow }{%
\triangledown }f_{so}\right) ^{2}(\lambda _{sc}-U)^{1/2}-\frac{1}{16}%
\triangledown ^{2}U(\lambda _{sc}-U)^{-1/2}\right]  \notag
\end{gather}%
This formula contains the "classical" Thomas-Fermi term plus an $\hbar ^{2%
\text{ }}$Wigner-Kirkwood correction. In this formula $U(\overrightarrow{r})$
is the central mean potential which contains also the Coulomb interaction
for the protons, and $f_{so}(\overrightarrow{r})$ is the spin-orbit
interaction. The classical turning points $r_{sc}$ are defined by $U(%
\overrightarrow{r_{sc}})=\lambda _{sc}$ where $\lambda _{sc}$ is the Fermi
level \newline
The numerical integration giving $N_{sc}$ is made with the help of an
"improved" Gauss-Legendre quadrature formula.\newline
The eigenvalues used in the Strutinsky level density are calculated by the
code published in \cite{25}.\newline
In the two methods we employ strictly the same Hamiltonian and the same set
of parameters, i.e. we use the Woods-Saxon potential with spin-orbit term
and the Coulomb potential for the protons. For this test we work with the $%
^{208}$Pb (neutrons) with a spherical nuclear shape. The parameters ( in $%
MeV,$ $fm$) are: $V_{0}=-47.083MeV$, $a_{v}=0.66MeV$, $R_{V}=7.36fm$, $%
\varkappa =12.0MeVfm^{2}$, $a_{so}=0.55fm$, $R_{so}=6.698fm$. Their
definition is given in \cite{25}.

Thus, in Fig.6, we have drawn the semiclassical level density (denoted by
semicl.) and the Strutinsky density as function of the Fermi level for three
cases (the numerical values of the smoothing parameters are given in the
figure).

\begin{itemize}
\item We can check in the three cases that the Strutinsky level density is
practically equal to semiclassical density in the "intermediate" region
(between the top and the bottom of the well) irrespectively of the order $M$
of the smoothing procedure. Indeed, we have shown in the subsection \ref{580}
that in the asymptotic limit (i.e. for medium and heavy nuclei) the
Strutinsky density of states (not the Strutinsky shell correction !) should
not be very sensitive to both free parameters of the method. However,
although with $M=0$ one obtains a good relative error on the density of
states, one improves more this error at sufficient large values of M.

\item As expected, due the reason invoked above, Fig.6 shows that the
Strutinsky densities differ from the semiclassical result essentially at the
top of the well. In the "intermediate region" there is no difference. Near
the singularity, it is more advantageous to choose high values for the order 
$M$ . With $M=30$, the Strutinsky calculation seem to work reasonably well
up to about $-2.5$ $MeV$, beyond this limit the precision is lost.

\item However, we should not forget that it is the semiclassical density
which is the "true" quantity. Due to the importance of the difference
between the two methods near the zero-energy, the semiclassical method must
in principle be preferred for the weakly bound nuclei.
\end{itemize}

\section{Conclusion:}

Although this paper explains a number of aspects and "subtleties" of the
Strutinsky method, we will insist on some essential points:

1) The Strutinsky level density appears in the fundamental Rel.(\ref{49})
only as an approximation ( and is not strictly equivalent as it is often
claimed) to the semiclassical level density. Consequently, the shell
correction calculated by the Strutinsky's method should also be considered
as an approximation to the semiclassical shell correction.

2) Semiclassical quantities such as the level density, the energy, the shell
correction must be considered as the "true" quantities compared to those
obtained with the Strutinsky method. Moreover, they do not depend on any
free parameter.

Unlike the semiclassical method, the Strutinsky method contains an intrinsic
noise (remainder). The ambiguity of the method comes from the dependence on
the two free parameters through this remainder.

3) It turns out that the remainder is proportional to $\left( \gamma
/\lambda \right) ^{M}$ and is defined in this paper as the relative error.
In the asymptotic limit $\left( \gamma /\lambda \right) \ll 1$, i.e. for
medium and heavy nuclei, the relative error is small, especially for higher $%
M$. Therefore, it is found that the Strutinsky method gives good results for
the average level density, and the average energy. In these cases the
dependency on the free parameters is weak.

On the contrary, in the shell correction the relative error is no longer
small. The shell correction might become strongly $\left( \gamma,M\right) $%
-dependent. The choice of these two free parameters must be treated with
care. In this context, in order to minimize the relative error, we propose
the "rule of the relative remainder (RRR)".

4) For realistic potentials, the semiclassical level density admits a
singularity at the top of the well. Since the Strutinsky method is a
least-squares approximation to the semiclassical level density (demonstrated
in this paper), the averaging fails near this singularity. In this case, the
two densities are different, and it is not surprising to note a strong
dependence on the parameters $(\gamma;M)$ in this region, even if the
continuum is treated properly. Consequently, for the weakly bound nuclei
(drip-line) it is better to use the semiclassical method.

Our personal conclusion, is that the semiclassical method with a "good
numerical treatment" should, in theory, be quite superior to the Strutinsky
method. The latter can be considered only as a good palliative method.

\appendix

\section{Calculation of the integral $I(M)$}

$I(M)=\dint \limits_{-\infty}^{\infty}{\small x}^{M+2}\widetilde{P_{M}}%
\left( x\right) \exp{\small (-x}^{2}{\small )dx}$, \ \ \ \ (see subsec.\ref%
{468})

First, one must note that $\widetilde{P_{M}}\left( x\right) $ of (\ref{4})
can be expressed as $\widetilde{P_{M}}\left( x\right) =\dfrac{H_{M}(0)}{%
2^{M+1}M!\sqrt{\pi}}\dfrac{H_{M+1}\left( x\right) }{x}.$

Indeed, with the help of the Christoffel-Darboux formula (chap 22 of \cite%
{16}):

$\dsum \limits_{k=0}^{n}\dfrac{H_{k}(x)H_{k}(y)}{2^{k}k!}=\dfrac{%
H_{n+1}(x)H_{n}(y)-H_{n+1}(y)H_{n}(x)}{2^{n+1}n!(x-y)}$

one finds for our case:

$\widetilde{P}_{M}\left( x\right) =\dsum\limits_{m=0}^{M}\dfrac{H_{m}(0)}{%
m!2^{m}\sqrt{\pi }}H_{m}\left( x\right) $

$=\dfrac{H_{M+1}\left( x\right) H_{M}(0)-H_{M+1}\left( 0\right) H_{M}(x)}{%
2^{M+1}M!\sqrt{\pi }(x-0)}$,

Since $M$ is even, $H_{M+1}\left( 0\right) =0$, this gives the cited
expression

To calculate the integral $I(M)$, we have to replace $\widetilde{P_{M}}%
\left( x\right) $ by the preceding result. We obtain:

$I(M)=\dfrac{H_{M}(0)}{2^{M+1}M!\sqrt{\pi}}\dint
\limits_{-\infty}^{\infty}x^{M+1}H_{M+1}\left( x\right) \exp(-x^{2})dx$

Now we use the following property \cite{16}:

$\dint \limits_{-\infty}^{\infty}t^{k}H_{k}\left( st\right) \exp(-t^{2})dt=%
\sqrt{\pi}k!P_{k}(s)$

where $P_{k}(s)$ is a Legendre polynomial. For our purpose, we choose $s=1$,
with $P_{k}(1)=1$. Making $k=M+1$ in the above result, one finds:

$\dint \limits_{-\infty}^{\infty}x^{M+1}H_{M+1}\left( x\right)
\exp(-x^{2})dx=\sqrt{\pi}\left( M+1\right) !$

so that:

$I(M)=\dfrac{H_{M}(0)}{2^{M+1}M!\sqrt{\pi}}\sqrt{\pi}\left( M+1\right) !=%
\dfrac{H_{M}(0)}{2^{M+1}}\left( M+1\right) $

Where $H_{M}(0)$ is given in the subsect.\ref{54}. Finally, the result can
be cast under the following form:

$I(M)=\dfrac{(-1)^{M/2}}{2^{\left( M+2\right) /2}}1.3.5......\left(
M+1\right) $

\section{Two applications of the Euler MacLaurin formula (EML)}

In the present work, to obtain some analytical results we employ the
Euler-MacLaurin formula (EML) \cite{16}.

$\dsum\limits_{k=1}^{n-1}F(k)=\dint\limits_{0}^{n}F(k)dk-\frac{1}{2}\left[
F(0)-F(n)\right] $

$+\frac{1}{12}\left[ F^{\prime }(n)-F^{\prime }(0)\right] -\frac{1}{720}%
\left[ F^{\prime \prime \prime }(n)-F^{\prime \prime \prime }(0)\right]
+.... $

Of course this formula can be used to calculate discrete finite sums. But
here, the interest of this formula is its application to the determination
of the asymptotic behavior $(n\rightarrow\infty)$ of the discrete sum. In
explicit terms, if we take a few terms (integral plus a few derivatives) in
this formula, we obtain a quantity which is equivalent to the discrete sum.
This means that the error (difference between the discrete sum and its
equivalent from the EML formula) tends to zero as $n$ increases to infinity.

In general, the higher orders of this formula are divergent and must be
simply ignored. For this reason, an asymptotic expansion does not exceed
three or four terms.

We apply this formula for two cases:

\begin{itemize}
\item The harmonic oscillator where we take just the integral in the EML
formula, neglecting thus the divergent terms (Dirac delta functions) in the
result:
\end{itemize}

$\dsum \limits_{n=0}^{\infty}D(n)$ $\delta\left(
\epsilon-(n+3/2)\hbar\omega\right) $

$\approx\dfrac{1}{\hbar\omega}\dint _{0}^{\infty}D(n)$ $\delta\left[ \dfrac{%
\epsilon}{\hbar\omega}-(n+3/2)\right] dn=g_{sc}(\epsilon)$

$D(n)=(n+1)(n+2)/2$ is the degeneracy of the level $n$, we find:

$g_{sc}(\epsilon )\approx \dfrac{1}{2\hbar \omega }\left[ \left( \dfrac{%
\epsilon }{\hbar \omega }\right) ^{2}-\dfrac{1}{4}\right] $

Now, it is easy to deduce the semiclassical energy:

$\overline{E}_{sc}(\lambda )\approx \dint\limits_{0}^{\lambda }\epsilon 
\overline{g}_{sc}(\epsilon )d\epsilon =\dfrac{\lambda ^{4}}{8\left( \hbar
\omega \right) ^{3}}-\dfrac{\lambda ^{2}}{16\hbar \omega }$

\begin{itemize}
\item In the cubic box the degeneracy is unknown, and we have to evaluate a
threefold integral. We work with a "basic" EML formula, i.e without
derivatives:
\end{itemize}

$\dsum \limits_{k=1}^{\infty}F(k)\approx\dint \limits_{0}^{\infty}F(k)dk-%
\frac{1}{2}F(0)$

Starting from that, we apply this formula in the case of the threefold sum:

While all axes are equivalent, we obtain

$\dsum\limits_{n_{x}=0}^{\infty }\dsum\limits_{n_{y}=0}^{\infty
}\dsum\limits_{n_{z}=0}^{\infty }F(n_{x},n_{y},n_{z})=$

$\dint\limits_{0}^{+\infty }\dint\limits_{0}^{+\infty
}\dint\limits_{0}^{+\infty }F(n_{x},n_{y},n_{z})dn_{x}dn_{y}dn_{z}$

$-(3/2)\dint\limits_{0}^{+\infty }\dint\limits_{0}^{+\infty
}F(n_{x},n_{y},0)dn_{x}dn_{y}$

$+(3/4)\dint\limits_{0}^{+\infty }F(n_{x},0,0)dn_{x}-(1/8)F(0,0,0)$

$=I_{3}+I_{2}+I_{1}+I_{0}$ \ \ \ (respectively)

Here $F(n_{x},n_{y},n_{z})=\delta\left(
\epsilon-(n_{x}^{2}+n_{x}^{2}+n_{x}^{2})E_{0}\right) $.

Using the spherical coordinates we find

$I_{3}=\dfrac{\pi}{4}\dfrac{\sqrt{\epsilon}}{E_{0}^{3/2}}$, \ \ \ \ \ $%
I_{2}=-\dfrac{3\pi}{8E_{0}^{{}}}$, \ \ \ \ \ $I_{1}=\dfrac{3}{8E_{0}^{1/2}%
\sqrt{\epsilon}}$, \ \ $I_{0}=-\dfrac{1}{8}\delta\left( \epsilon\right) $

Here also we omit $I_{0}$( delta function)

$g_{sc}(\epsilon )=\dfrac{\pi }{4}\dfrac{\sqrt{\epsilon }}{E_{0}^{3/2}}-%
\dfrac{3\pi }{8E_{0}^{{}}}+\dfrac{3}{8E_{0}^{1/2}\sqrt{\epsilon }}$

Therefore:

$\overline{E}_{sc}(\lambda )\approx \dint\limits_{0}^{\lambda }\epsilon 
\overline{g}_{sc}(\epsilon )d\epsilon =\dfrac{\pi }{10}\dfrac{1}{E_{0}^{3/2}}%
\lambda ^{5/2}-\dfrac{3\pi }{16E_{0}^{{}}}\lambda ^{2}+\dfrac{1}{4E_{0}^{1/2}%
}\lambda ^{3/2}$

\end{document}